\documentclass{elsart}

\usepackage{amssymb}
\usepackage[all]{xy}
\begin{document}
\begin{frontmatter}

\title{RDDI phase in hydrogenated metals -some clarifications}
\author{J.S.Brown}
\address{Clarendon Laboratory, University of Oxford, Parks Rd. OX1 3PU, UK}

\begin{abstract}
We present further evidence for the coherent RDDI phase postulated in our earlier paper and clarify the reasons why the intersite tunnelling amplitude is increased by many orders of magnitude when a mesoscopic domain undergoes such a transition.
\end{abstract}

\end{frontmatter}

\section{Remarks and clarifications}
\label{rac}
In view of the wide interest shown in our previous paper \cite{Bro06}, we consider it appropriate to clarify a few issues that were inadequately dealt with there. Firstly, we would like to thank Prof. Kuritzi for having drawn our attention to earlier work \cite{Kur96} on the exchange of single-mode photons between spatially separated independent two-level systems. Since this mechanism was termed "resonant dipole-dipole interaction" (RDDI), we will forthwith use this in place of our provisional "coherent dipole-dipole oscillation".
\\
 Secondly, we would like to draw attention to studies of KDP \cite{Tok85} and its deuterated analogue that we consider to provide direct evidence of a low temperature RDDI phase in an {\it ionic} crystal, thereby emphasizing the generic nature of the effect. The observed transition temperatures reported were 122 K for protons and 220 K for deuterons. The neutron TOF spectroscopy data are not consistent with conventional explanations in terms of optic phonon modes, but appear to be entirely commensurate with an intercell dipole-dipole interaction at an essentially single, temperature independent, frequency.
\\
Thirdly, we would like to make it clear that we are well aware that our model of identical cells exchanging monochromatic transverse photons at $T=0$ is a gross approximation. In a real crystal, there will be line-broadening due to i) the photon-phonon coupling, ii) thermal photons, iii) proximity to the surface region and iv) incomplete hydrogenation. To what extent our neglect of these and other effects would mitigate against the actual manifestion of the proposed RDDI phase remains to be seen.
\\
Finally, we would like to justify our somewhat vague allusions to enhancement of the D-D fusion rate in metals due to RDDI phase onset by means of the following qualitative argument.
\\
Denoting the the ground (+++ parity) state of the D adatom by $s$ and the excited (-++ parity) member of the first excited triplet by $p$, it is clear that, if the intersite dipole-dipole interaction is large enough compared to the excitation energy, the new RDDI ground state $\Psi$ is a superposition of states like $|..spspspspspspspspsp..>$ and permutations thereof obtained by exchanging an $s$ with a neighbouring $p$.
\\
In view of the short range of the dipole-dipole interaction and the consequent sparseness of the interaction matrix, $\Psi$ will be dominated by contributions from states differing from the alternating state $|..spspspspspspspspsp..>$  by just a few $s-p$ exchanges.
Since each one of these states is a degenerate solution of the many-body Hamiltonian of energy $N(\epsilon_s + \epsilon_p)/2$, the resulting superposition gives rise to coherent interference in the adatom coordinate space of dimension $3N$. Following the destructive interference of nearly all configurations, the very few remaining reinforced configurations correspond to a collective classical motion of all D adatoms along the dipole direction. In a 1D-chain, for instance, adjacent neighbours will appear to move in opposite directions. 
\\
This is best illustrated by a consideration of the simplest of all possible cases: a 1D-chain of $N=2$ occupied cells and a coarse mesh of just two possible positions {01,10}, one on each side (along the dipole and chain axis) of the cell. Of the four configurations 10:10, 10:01, 01:10, 01:01, equally likely in the incoherent regime, only 10:01 and 01:10 survive in the RDDI ground state $\Psi = 2^{-0.5}( |sp>-|ps> )$.
\\
It is hence easy to see that in a 1D-chain of arbitrary length, the following two configurations have the largest amplitude in the RDDI ground state:
\\
01:10:01:10:01:10:01:10........01:10:01:10:01:10  and
\\
10:01:10:01:10:01:10:01........10:01:10:01:10:01
\\
The normalized amplitude of these dominant configurations is on the order of $2^N$ times greater than in the normal incoherent regime, all cross-terms vanishing by virtue of the orthogonality of the component states. The probability that any one adjacent pair at 01:10 have tunneled through the classically forbidden region under their mutual Coulomb barrier is accordingly multiplied by the same exponential factor (N.B. the tunnelling probability is proportional to the square of the sum of very many, extremely small, {\it unipolar} contributions, multiplied by the oscillation frequency). In a mesoscopic region comprising many hundreds of adatoms, this factor amounts to many orders of magnitude and may transform the otherwise vanishingly small fusion rate into an experimentally observable phenomenon with technological potential.
\\
In view of the finite rate of particle exchange in the bridging sites, the state of N coherent bosonic deuteron adatoms will quickly become exchange-symmetric. Because of this, the amplitude of any one D-D fusion event will be shared equally over all sites. This translational symmetry will presumably forbid the emission of quanta of wavelength small compared to the coherence domain and force a relatively slow radiationless relaxation of the fused deuterons to helium-4.

\end{document}